# IG-CFAT: An Improved GAN-Based Framework for Effectively Exploiting Transformers in Real-World Image Super-Resolution

Alireza Aghelan*[1], Ali Amiryan[2], Abolfazl Zarghani[1], Modjtaba Rouhani[1]

[1]Department of Computer Engineering, Ferdowsi University of Mashhad, Mashhad, Iran
[2]Department of Computer Engineering, Amirkabir University of Technology, Tehran, Iran
aghelan.alireza@mail.um.ac.ir, ali.amiryan@mail.aut.ac.ir, abolfazlzarghani1999@mail.um.ac.ir, rouhani@um.ac.ir

## Abstract

In the field of single image super-resolution (SISR), transformer-based models, have demonstrated significant advancements. However, the potential and efficiency of these models in applied fields such as real-world image super-resolution have been less noticed and there are substantial opportunities for improvement. Recently, composite fusion attention transformer (CFAT), outperformed previous state-of-the-art (SOTA) models in classic image super-resolution. In this paper, we propose a novel GAN-based framework by incorporating the CFAT model to effectively exploit the performance of transformers in real-world image super-resolution. In our proposed approach, we integrate a semantic-aware discriminator to reconstruct fine details more accurately and employ an adaptive degradation model to better simulate real-world degradations. Moreover, we introduce a new combination of loss functions by adding wavelet loss to loss functions of GAN-based models to better recover high-frequency details. Empirical results demonstrate that IG-CFAT significantly outperforms existing SOTA models in both quantitative and qualitative metrics. Our proposed model revolutionizes the field of real-world image super-resolution and demonstrates substantially better performance in recovering fine details and generating realistic textures. The introduction of IG-CFAT offers a robust and adaptable solution for real-world image super-resolution tasks. The source code is available at https://github.com/alireza-aghelan/IG-CFAT

# 1. Introduction

Image super-resolution is the process of generating high-quality images from corresponding low-resolution versions. This technique is useful in medical imaging, satellite observation, underwater exploration, and various multimedia applications. With the emergence of deep learning, image super-resolution models employed basic architectures, such as convolutional neural networks (CNN), and produced a high Peak Signal-to-Noise Ratio (PSNR). However, these models tend to create an image that is good in quantitative metrics but lacks the fine details that make them visually appealing.

These limitations are overcome by recent approaches including generative adversarial networks [1] (GANs) – based and transformer-based methods. The GAN-based techniques advance textural realism through adversarial training: pitting a generative network against a discriminative one and learning to produce significantly more detailed and visually attractive images. Some recent works have presented transformer-based models, like SwinIR [2] and hybrid attention transformer (HAT) [3], making significant advancements in this area. These models leverage the power of the transformers to capture long-range dependencies in the models and enhance global consistency and fine details of super-resolved outputs. However, classic GAN-based and transformer-based models cannot effectively enhance the perceptual quality of images degraded by complex and unknown degradation patterns common in real-world scenarios.

Over time, the use of super-resolution (SR) methods has been gaining more and more importance in real-world scenarios since they offer a fundamental breakthrough in visual quality. Real-world image super-resolution acknowledges the fact that image degradations are complex, most of them being unpredictable and without a proper representation in the available controlled datasets. Models, such as BSRGAN [4] and Real-ESRGAN [5] have partially addressed this problem by using real-world degradation models, though they still have some drawbacks. However, recent advanced degradation models have shown superior performance in mimicking real-world degradations, resulting in significantly improved perceptual image quality.

Most of the GAN-based super-resolution models that use transformer-based generators, such as SwinIR-GAN and Real-HAT-GAN [6], have used conventional degradation models, which may restrict the model's effectiveness in adapting to unpredictable degradation patterns in real-world scenarios. Moreover, these models utilize relatively basic discriminators that do not consider the semantics of images, resulting in outputs that lack realistic texture and accurate detail. Therefore, to fully unleash the potential of transformers in real-world image super-resolution tasks, fundamental modifications, including the use of more advanced degradation models and semantic-aware discriminators, are necessary. Furthermore, GAN-based super-resolution models typically use a combination of L1 loss, perceptual loss, and GAN loss functions for training. However, these loss functions often have limitations in capturing high-frequency details. As a result, incorporating a specialized loss function to effectively capture high-frequency information is essential.

Among recent transformer-based super-resolution models, the Composite Fusion Attention Transformer [7] (CFAT) stands out as a novel model that utilizes non-overlapping triangular windows alongside traditional rectangular windows. This technique enables the model to access more unique shifting modes and reduces boundary distortions, resulting in improved performance. This paper proposes a novel GAN-based framework incorporating the CFAT model to advance the applicability and effectiveness of real-world image super-resolution models. In this work, we aim to effectively exploit the power of transformers for real-world image super-resolution by applying several improvements including a semantic-aware discriminator, an adaptive degradation model, and a wavelet loss function.

By incorporating a semantic-aware discriminator [8], IG-CFAT can better enhance texture and detail quality by leveraging deep semantic understanding from pre-trained vision models. Moreover, employing an adaptive degradation model [9,10] improves the robustness and adaptability of the model toward complex degradation patterns in real-world scenarios. Furthermore, the addition of wavelet loss to the loss functions of GAN-based models addresses the shortcomings of these models in reconstructing high-frequency details. Extensive experiments demonstrate that IG-CFAT significantly outperforms previous GAN-based real-world image super-resolution models in quantitative and qualitative metrics. Our model excels in recovering fine details and generating realistic textures, resulting in superior fidelity and perceptual quality. These improvements make our model a robust and adaptable solution for real-world applications.

The main contribution of our paper includes:

1- We introduce a novel GAN-based framework named IG-CFAT to effectively harness the performance of transformers for real-world image super-resolution.
2- In our framework, we integrate recent advancements and techniques including a semantic-aware discriminator [8] to better recover textural detail and an adaptive degradation model [9,10] to improve the simulation of complex degradations.
3- We propose a new integration of loss functions by adding wavelet loss to conventional loss functions of GAN-based super-resolution models to reconstruct high-frequency details more efficiently.
4- Comprehensive experiments show that IG-CFAT significantly outperforms GAN-based super-resolution models by effectively recovering fine details and generating realistic textures, resulting in superior quantitative and qualitative metrics.

The rest of this paper is divided into four sections. In Section 2 we explore the research highlighting advancements in real-world image super-resolution. Section 3 describes the materials and methods covering the degradation process, model architecture, and loss functions. Section 4 provides an analysis discussing the datasets used, implementation details, and comparing the IG-CFAT model with state-of-the-art methods. Finally, Section 5 wraps up the paper, by summarizing our contributions and exploring future directions, for research.

# 2. Related Work

The related work is divided into three sections. First, we provide an overview of real-world image super-resolution methods. In the subsequent sections, we explore degradation models and discriminators in more detail.

## 2.1. Real-World Image Super-Resolution Methods

Classic real-world image super-resolution methods often use bicubic downsampling and traditional degradation models. The simplicity of their degradation models limits their applicability as these methods are unable to adapt to scenarios where the degradation patterns are unpredictable. GAN-based models such as BSRGAN [4] and Real-ESRGAN [5] have brought progress to the field of real-world image super-resolution by incorporating practical degradation models. Recent research has further improved upon these advancements by proposing adaptive degradation models. One notable example is the adaptive degradation model in the DASR [9] model, which enables effective image restoration especially when dealing with unpredictable degradations.

SwinIR-GAN [2] and Real-HAT-GAN [6] are great examples of GAN-based models that employ transformer-based generators. These models benefit from the advanced characteristics of transformers, however, they haven't adopted the most advanced degradation models. SwinIR-GAN used the degradation model of the BSRGAN and Real-HAT-GAN employed the high-order degradation model of the Real-ESRGAN. Additionally, these models rely on relatively simple discriminators that ignore the semantic content of images, leading to outputs with less realistic textures and precise details. This suggests a need for semantic-aware discriminators to effectively utilize the strengths of transformers in real-world image super-resolution.

In most GAN-based super-resolution models, a typical strategy includes the use of a combination of L1 loss, perceptual loss, and GAN loss. L1 loss minimizes the average absolute differences between the predicted and true pixel values to increase overall fidelity.

Moreover, perceptual loss encourages perceptually relevant characteristics by aligning features extracted from pre-trained networks. GAN loss helps to produce images that are indistinguishable from real images and improves textural realism. However, these conventional loss functions often fail to capture high-frequency details, crucial for visual quality. To overcome this, a specialized loss function focusing on high-frequency information is essential.

In our work, we aim to address all aforementioned limitations to produce a comprehensive and practical real-world image super-resolution model.

## 2.2. Degradation Models

Most of the classic degradation models mainly simulate image degradation by applying a predefined sequence of blur, downsampling, noise, and JPEG compression. In many cases, such models cannot fully simulate the complex and diverse degradations in the real world, and this has caused suboptimal performance in real-world applications often leading to output images that lack realistic textures and accurate details.

To address this weakness of classic degradation models, real-world degradation models were developed. For example, BSRGAN uses a sophisticated degradation model involving random shuffle blur, downsampling, and noise degradations. This is one of the first effective degradation models for handling real-world degradations. Similarly, Real-ESRGAN adopted a high-order degradation modeling process that more accurately represents complex real-world degradations. In the high-order degradation, common artifacts such as ringing and overshoot, which are often overlooked in simpler models, are included. Improving these aspects, Real-ESRGAN will make the super-resolution process more robust and more effective, hence more practical for use in image restoration. To further enhance the applicability and effectiveness of real-world image super-resolution models, Liang et al. proposed the DASR model. This model utilizes an innovative degradation-adaptive super-resolution network that dynamically adjusts its parameters based on the estimated degradation of each input image. Such forms of adaptability enable

DASR to effectively handle images with varying levels of degradation, making it highly effective for real-world applications.

To improve the effectiveness of our model in handling real-world degradations, we incorporate an adaptive degradation model inspired by DASR and StarSRGAN [10].

## 2.3. Discriminator Networks

Recently, some work has focused on improving the discriminator networks in GAN-based models for super-resolution. In this section, we discuss some critically influential improvements that have greatly enhanced the performance of discriminators.

In the ESRGAN [11] model, a relativistic discriminator [12] predicts relative realness instead of the absolute value, resulting in improved texture detail and realism of the outputs. Park et al. [13] introduced a GAN-based super-resolution framework having two discriminators; one of them works in the image domain, while the other one performs discrimination in the feature domain. The feature discriminator has dramatically improved the ability of the generator to produce structurally detailed high-frequency features without artifacts, resulting in improved perceptual quality. Real-ESRGAN adopted a U-Net-based discriminator [14] that integrates spectral normalization [15] to enhance local texture feedback while preserving global style assessment. Spectral normalization stabilizes training dynamics and reduces artifacts. This discriminator enables a delicate balance between detail enhancement and artifact suppression. A-ESRGAN [16] uses a multi-scale attention U-Net-based discriminator. The attention mechanism enables better focusing on the edges, resulting in sharper details with reduced distortion. Moreover, the multiscale strategy enhances the ability of the model to reconstruct fine textures. Despite these advancements, the aforementioned discriminators often overlook the semantics of the image, which may result in outputs that lack realistic textures and accurate details.

In our IG-CFAT model, we harness the semantic-aware discriminator [8] (SeD) which substantially surpasses previous approaches.

# 3. Materials and methods

Our approach aims to effectively exploit the capabilities of transformers for real-world image super-resolution. To this end, we have utilized three key components: an adaptive degradation model, a semantic-aware discriminator network, and a wavelet loss function. To explain the methodology, we begin by describing the degradation model. This is followed by an in-depth look at the generator and discriminator networks. Finally, we discuss the wavelet loss function employed to better reconstruct the high-frequency details of the input images.

## 3.1. Degradation Model

Most of the recent GAN-based models have adopted the high-order degradation model of the Real-ESRGAN [5]. This model aims to simulate real-world image degradations by applying the same degradation process multiple times. This method is effective, however, it may limit the model's ability to adapt to the unpredictable degradation patterns commonly found in real-world scenarios. To overcome this problem, the IG-CFAT uses an adaptive degradation model [9,10].

The adaptive degradation model has significant merits compared to the high-order degradation model. This degradation model can capture a wider range of degradation patterns, making it more practical for real-world scenarios. This model segments the degradation space into three levels: weak degradation (D1), standard degradation (D2), and severe degradation (D3). Levels D1 and D2 are first-order degradation with small and large parameter ranges, respectively. Level D3 is second-order degradation. One of these degradation levels is randomly selected to generate LR-HR image pairs. The probability distribution for the selection of degradation levels is [0.3, 0.3, 0.4]. To better understand the degradation model's performance, we illustrate its working method in Figure 1. We have depicted the parameters of blur, noise, downsampling, and JPEG compression at each stage to show the differences between the three degradation levels.

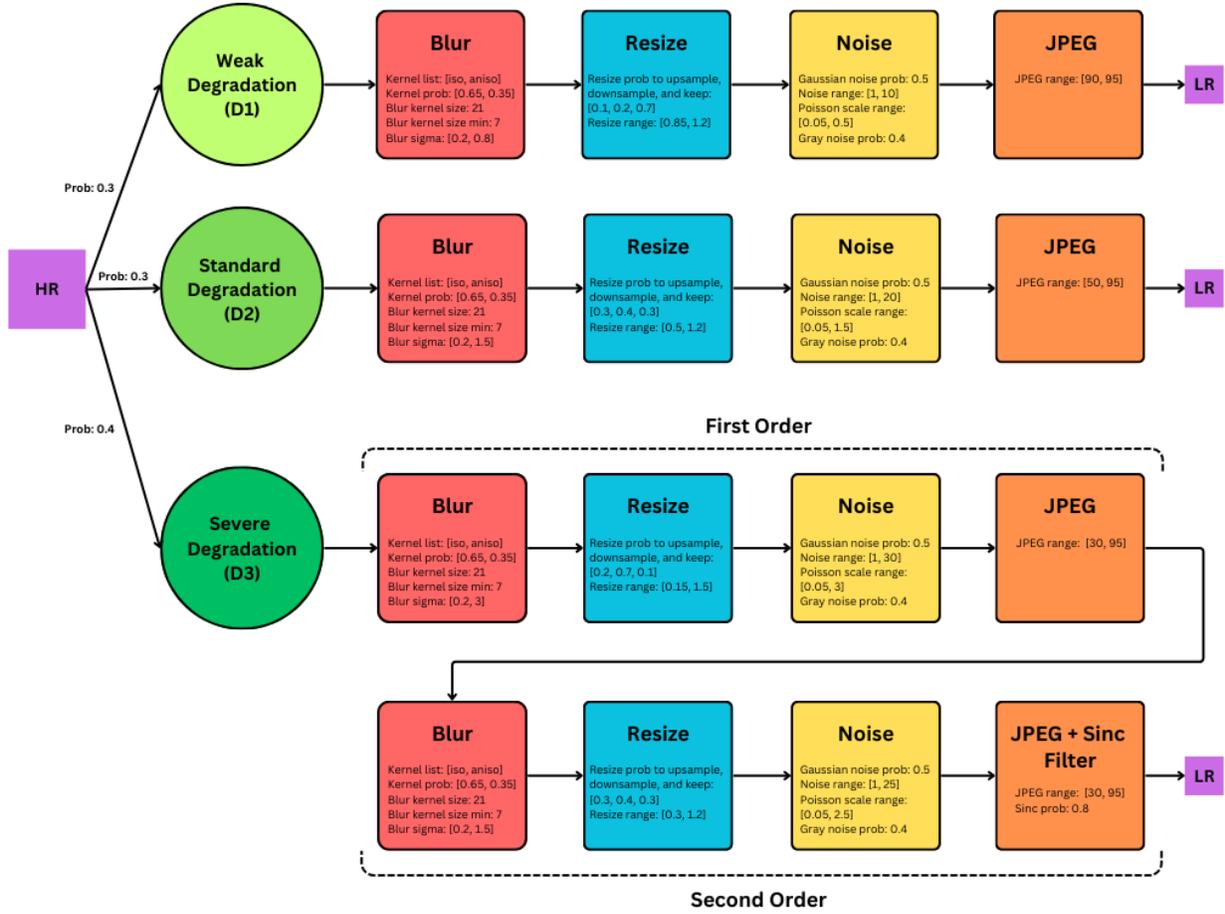

**Figure 1.** Diagram of the adaptive degradation model.

## 3.2. Networks Architecture

### 3.2.1. Generator Network

The overall architecture of the IG-CFAT is shown in Figure 2. The generator is the same as in the original CFAT model [7]. Previous transformer-based models, which rely on overlapping rectangular shifted windows, often suffer from boundary distortion and limited unique shifting modes. To address these problems, CFAT incorporates a novel non-overlapping triangular window technique that works simultaneously with rectangular windows. This method can reduce distortions at the boundaries and give it access to more varieties in shifting modes. Moreover, CFAT combines triangular-rectangular window-based local attention with a channel-based global attention technique to further improve performance. As a result, the CFAT leverages the power of transformers more effectively, making it an ideal choice to be used as a generator within our framework.

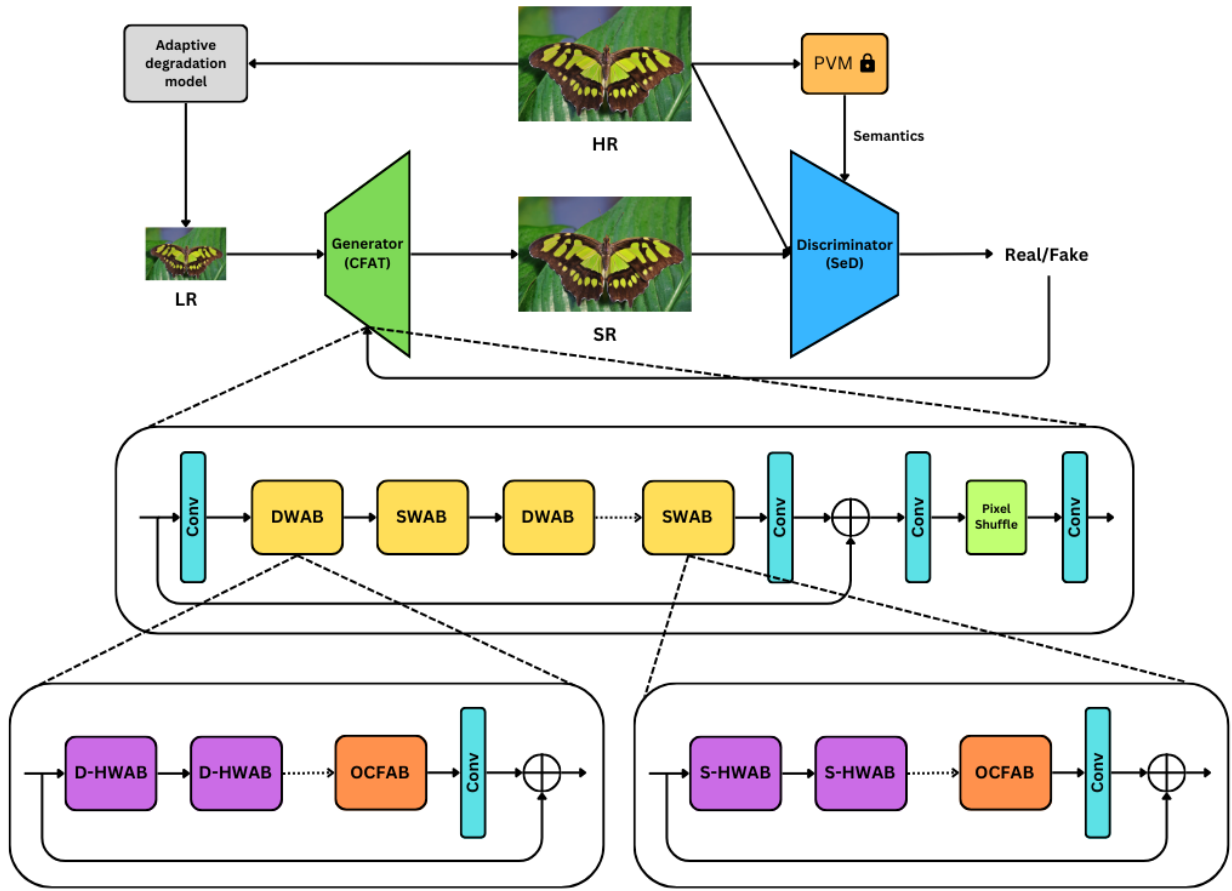

**Figure 2.** The overall architecture of IG-CFAT.

The CFAT model has a complex architecture, including Dense Window Attention Blocks (DWAB) and Sparse Window Attention Blocks (SWAB). These blocks utilize both rectangular and triangular window attention units for detailed feature extraction. The Overlapping Cross Fusion Attention Block (OCFAB) enhances the performance by overlapping features across neighboring windows and establishing cross-attention between them, using a sliding window technique. Moreover, Channel-Wise Attention Block

(CWAB) utilizes depthwise-pointwise convolutions to reduce the squeeze factor and improve performance. The CFAT architecture is shown in Figures 2 and 3. More details about CFAT architecture can be found in [7].

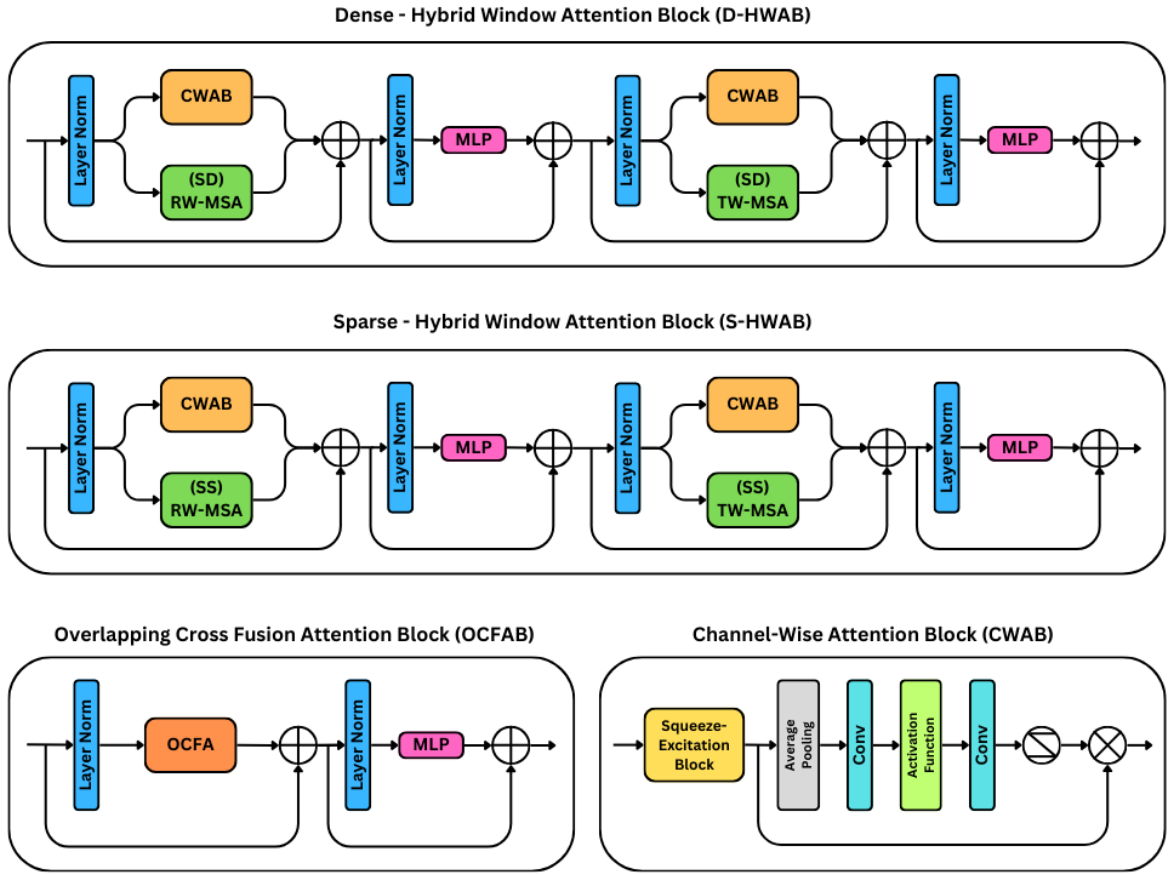

**Figure 3.** Internal architecture of some blocks, including D-HWAB, S-HWAB, OCFAB, and CWAB.

### 3.2.2. Discriminator Network

Discriminators in conventional GAN-based models for image super-resolution mainly evaluate the general authenticity of super-resolved images by measuring the distribution distance between generated and reference images. This method is effective in distinguishing real images from fake ones, however, it primarily evaluates texture and detail at a coarse-grained level. Most discriminators of this kind ignore the semantics of the image. This approach may lead to outputs without realistic textures and accurate details.

To overcome these limitations, our model employs a semantic-aware discriminator [8] (SeD). Different from previous models, SeD utilizes the semantics of images as a condition; hence, it can perform fine-grained analysis of the texture and details. This is achieved by extracting semantic information from the middle features of widely used pre-trained vision models (PVMs). In this way, the discriminator evaluates the authenticity of textures and respects the semantics of the image, resulting in more precise and context-aware texture generation.

In [8], the authors incorporated SeD into two popular discriminators, including a patch-wise discriminator and a pixel-wise discriminator. In our work, we utilize the pixel-wise semantic-aware discriminator integrated with a U-Net architecture (SeD + U). The SeD improves texture realism by leveraging image semantics, leading to more photo-realistic and pleasing super-resolved images. These improvements make the SeD a crucial component in our IG-CFAT model, significantly contributing to better performance in real-world image super-resolution tasks.

## 3.3. Loss Functions

As mentioned in Section 2.1, in most GAN-based super-resolution models, a combination of loss functions, including L1 loss, perceptual loss, and GAN loss, is employed, but they often fail to capture high-frequency details. To address this limitation and further capture high-frequency details essential for visually pleasing results, we present a novel combination of loss functions by integrating wavelet loss with the conventional loss functions of the GAN-based super-resolution models. Inspired by [17], the images are first converted from the RGB color space to the YCbCr color space. Next, the Stationary Wavelet Transform (SWT) is used to decompose the Y channel into its frequency subbands. The SWT decomposes the Y channel of the image into one low-frequency (LF) subband, called LL, and multiple high-frequency (HF) subbands, named LH, HL, and HH. Wavelet loss function implementation is as follows:

$$L_{SWT} = \mathbb{E}\left[\sum_j \lambda_j \|SWT(G(x))_j - SWT(y)_j\|_1\right]$$

where G denotes the super-resolution model, λ j are scaling factors that balance the importance of each subband, x is the LR image, and y is the ground truth image. The loss is calculated by evaluating the L1 fidelity loss between the SWT subbands of the generated images and the ground truth images. Figure 4 illustrates how the wavelet loss is computed.

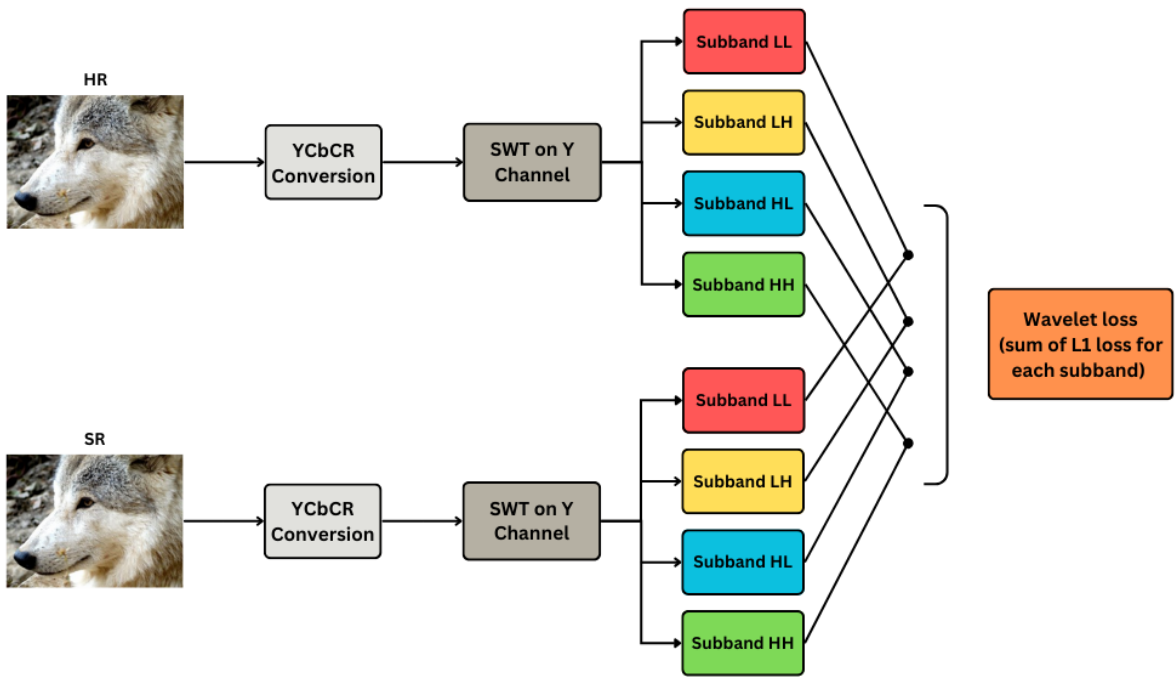

**Figure 4.** Wavelet loss computation diagram.

To determine the total loss of the generator, we add wavelet loss to the L1 loss, perceptual loss, and GAN loss. The total loss of the generator can be represented as:

$$L_{\text{total}} = L_1 + L_{\text{perceptual}} + L_{\text{SWT}} + L_{\text{GAN}}$$

Conventional losses in GAN-based models often fail to effectively capture high-frequency details. The incorporation of wavelet loss addresses this shortcoming by improving the ability of the model to reconstruct high-frequency details. Therefore, the model achieves superior fidelity (higher PSNR) and boosts the perceptual quality of the output images.

## 4. Experiments

### 4.1. Datasets

To train our model, we use the DIV2K [18] and Flickr2K [19] datasets. DIV2K includes 800 images, and Flickr2K provides 2,650 images. We utilize original images from the DIV2K and Flickr2K datasets without employing any data augmentation techniques. This resulted in a total of 3450 training images. For testing our model, we use the DIV2K validation set. We exclude one image from the DIV2K validation set as its height of 816 pixels falls below our threshold of 1000 pixels. We generate ground truth images by resizing original images to either 2000x1000 or 1000x2000 using Lanczos interpolation. These ground truth images are then downsampled by a factor of four using Bicubic interpolation to produce their LR counterparts.

### 4.2. Implementation Details

We first train our generator with L1 loss and adaptive degradation model for 86300 iterations, equal to 100 epochs. Then the pre-trained generator is integrated into our GAN framework and fine-tuned for 43150 iterations, equal to 50 epochs. In both stages, we use the Adam optimizer and maintain a learning rate of 0.0001. To improve perceptual quality, we utilize the unsharp masking (USM) technique. This process, including training and fine-tuning phases, is executed on an RTX 3090 GPU, with a batch size of 4.

## 4.3. Comparison with State-of-the-Art Methods

To evaluate the performance of our proposed model, we conduct extensive comparisons with SOTA real-world image super-resolution models including, BSRGAN, Real-ESRGAN (x4plus version), A-ESRGAN (x4plus version - trained with multi-scale attention U-Net-based discriminator), SwinIR-GAN (large version), and Real-HAT-GAN (sharper version). We use the officially released versions of these models for comparison.

### 4.3.1. Quantitative Results

We use Peak Signal-to-Noise Ratio (PSNR) and Learned Perceptual Image Patch Similarity (LPIPS) metrics to evaluate the performance of the models. The quantitative results are summarized in Table 1. Moreover, the training datasets used for each model are detailed in Table 1. The quantitative assessment clearly illustrates the superior performance of the IG-CFAT model. With a PSNR of 26.81, IG-CFAT surpasses all compared models in fidelity. In terms of the LPIPS metric, our model also leads with the lowest score of 0.175, demonstrating the superior perceptual quality of its outputs.

| Model | Training Datasets | PSNR | LPIPS |
|---|---|---|---|
| BSRGAN | DF2K + WED + …* | 24.85 | 0.229 |
| Real-ESRGAN | DF2K + OST | 24.39 | 0.222 |
| A-ESRGAN | DIV2K | 22.75 | 0.274 |
| SwinIR-GAN | DF2K + OST + …* | 24.72 | 0.206 |
| Real-HAT-GAN | DF2K + OST | 25.02 | 0.203 |
| **IG-CFAT** | DF2K | 26.81 | 0.175 |

**Table 1.** Quantitative comparison between IG-CFAT and GAN-based SR methods. The best and second performances are marked in red and blue. * More details about the datasets used for training can be found in the reference paper.

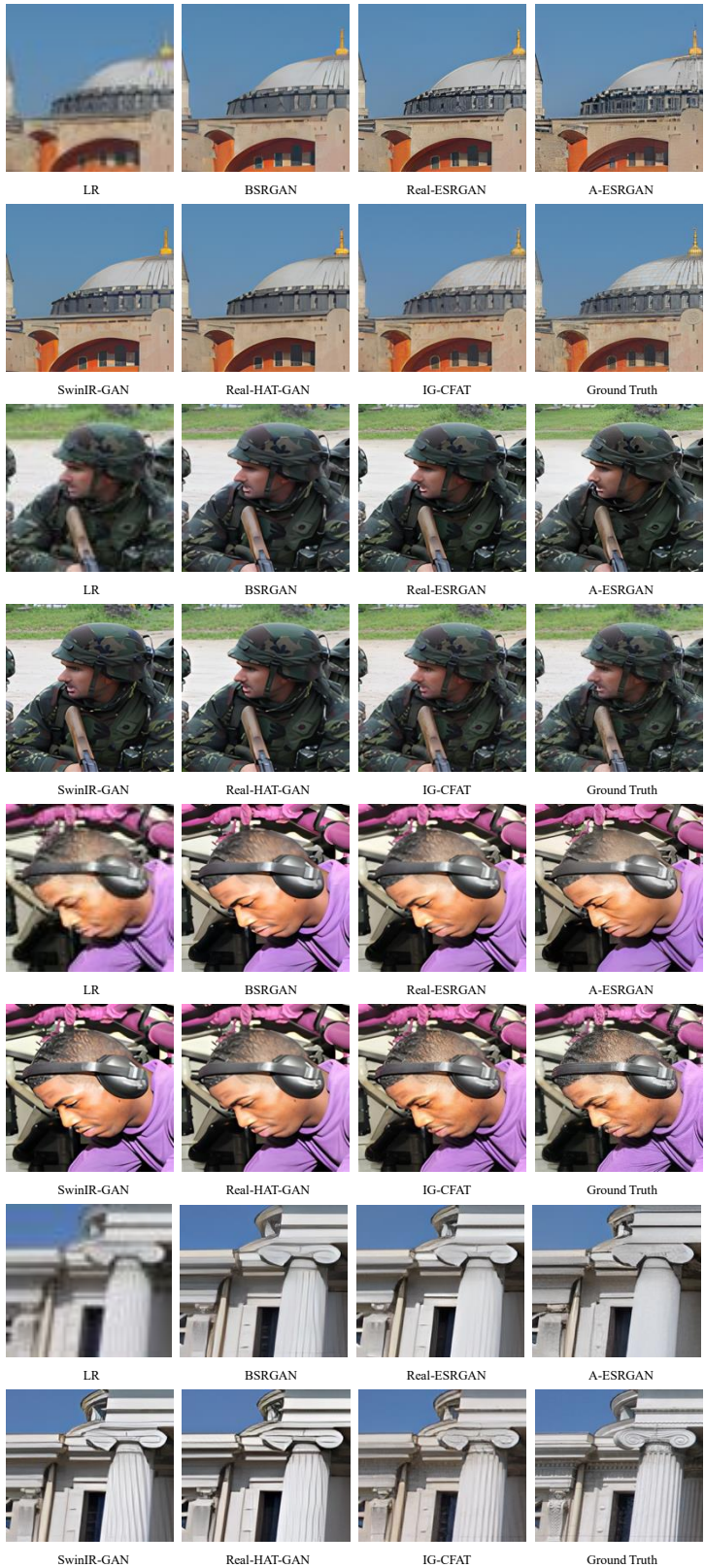

**Figure 5.** Visual comparison of IG-CFAT with GAN-based SR methods. (Zoom in for a better view)

### 4.3.2. Qualitative Results

The qualitative results are presented in Figure 5. They illustrate the superior performance of the IG-CFAT model in reconstructing fine details and generating realistic and natural textures. In the first picture, the model accurately recovers the designs and textures of the dome, and in the last image, IG-CFAT effectively restores the column's details and textures and makes it closely resemble the original structure. Moreover, the images restored by IG-CFAT exhibit enhanced clarity. In the second and third images, IG-CFAT renders the soldier's and the man's facial features with exceptional clarity and detail.

Quantitative and qualitative results demonstrate that IG-CFAT sets new benchmarks in the field of real-world image super-resolution. The combination of the robust and adaptable performance of the IG-CFAT model makes it a promising solution for practical applications that require high-quality images, including medical imaging, underwater imaging, surveillance, and consumer multimedia. For employing real-world image super-resolution models in specific fields, it is advisable to fine-tune them on targeted datasets. For example, fine-tuning real-world image super-resolution models on medical image datasets improves results as shown in [20], and similar benefits are seen with underwater images as demonstrated in [21].

## 5. Conclusion and Future Directions

This paper presented IG-CFAT, an improved GAN-based model to effectively exploit the performance of transformers for real-world image super-resolution. We employed a semantic-aware discriminator that evaluates the realism of super-resolved images and their semantic coherence, resulting in more accurate and perceptually pleasing outputs. Furthermore, the adaptive degradation model improves the model's robustness and generalization capabilities in handling real-world degradations. Moreover, using wavelet loss enables the model to better reconstruct high-frequency details and generate higher perceptual quality images. By incorporating these advancements, IG-CFAT significantly outperforms previous SOTA models in both quantitative and qualitative metrics.

One of the future directions is to further refinements and extensions of the degradation model by exploring the task grouping based real-SR method (TGSR) [22] to broaden the applicability of real-world image super-resolution models. Moreover, employing the patch-wise semantic-aware discriminator (SeD + P) may enhance the performance of the model. In addition, using the LSDIR [23] dataset can improve performance and generalization capabilities in image restoration tasks. The LSDIR dataset is a large-scale collection of high-resolution images, specifically prepared for image restoration tasks.